\documentclass[journal=jacsat,manuscript=letter,layout=twocolumn]{achemso}

\usepackage{chemformula} 
\usepackage[T1]{fontenc} 
\usepackage{braket}
\usepackage{xcolor}
\usepackage{abstract}
\usepackage{cuted}
\usepackage{etoolbox}

\author{Maciej Hendzel}
\affiliation[Jagiellonian University]
{Institute of Theoretical Physics, Jagiellonian University, ulica \L{}ojasiewicza 11,\\ PL-30-348 Krak\'{o}w, Poland}
\author{Maciej Fidrysiak}
\author{J\'ozef Spa\l{}ek}
\email{jozef.spalek@uj.edu.pl}
\title[Towards Complementary Characterization \\ of the Chemical Bond]
  {Towards Complementary Characterization \\ of the Chemical Bond}

\begin{document}

\begin{strip}
\vspace{-0.5cm}
\begin{abstract}
A precise discussion of a single bond requires consideration of two-particle wave function for the particles involved. Here we define and determine rigorously the intrinsic covalency and connected characteristics on the canonical example of \ch{H2} molecule. This is achieved by starting from analytic form for the two--particle wave function for electrons forming the bond, in which we single out the atomic contribution (\textit{atomicity}) in an unequivocal manner. The presence the of atomicity and ionicity factors complements the existing attributes of the bond. In this way, a gradual evolution of the molecular state to its two-atomic correspondant is traced systematically with increasing interatomic distance. In effect, a direct relation to the onset of incipient
Mott-Hubbard atomicity (\textit{Mottness}) to the intrinsic covalency and ionicity is established. This goal is achieved by combining the single--particle wave function readjustment with a simultaneous determination of two--particle states in the particle (second--quantization) representation.
\end{abstract}
\end{strip}

The concept of chemical bond and its quantum properties are of fundamental importance to our understanding of both physical and chemical characteristics of molecules and solid state compounds \cite{Szabo1987, Piela2013, Bacskay2017}. Among the principal questions are those of relative role of covalency, ionicity, and atomicity, as they describe qualitative differences with the characteristics of parent atomic states composing the system. Here we propose a resolution of the question concerning the evolution of the molecular into corresponding atomic states and vice versa. In effect, such an approach leads to an unequivocal determination of true covalency, as well as to extracting both the atomicity and ionicity. As a side result, we resolve the longstanding question of unphysical behavior of covalency with the increasing interatomic distance (see also Supporting Information).  

At the outset, we take a multiparticle view of the chemical bond and implement a special method EDABI (\textbf{E}xact \textbf{D}iagonalization \textit{\textbf{Ab I}nitio} approach) devised in our group earlier \cite{JS2000, JPCM2007, SpalJul} and apply it here to a rigorous analysis in the simplest situation of the \ch{H2} molecule. The generic case of the \ch{H2} molecule involves two--electron single bond and thus the role of interelectronic correlations in combination with concomitant single-particle wave function readjustment in the resultant (correlated) state, can be analyzed rigorously and in mathematically analytic terms. 

The structure of this Letter is as follows. After introducing exact two-particle electron wave function we redefine the covalency to extract from standard definition the contribution of atomicity, by referring to the notions of Mott and Hubbard localization. This new formulation allows us to define and determine explicitly the true covalency, atomicity (also termed seniority \cite{ChenSeniority}), and ionicity factors in the chemical bond and hence to understand those component factors of the chemical bond in a fully quantitative manner. The whole methodology is based on combination of both the first-- and second--quantization aspects of the relevant multielectron states, detailed in the Supporting Information. 

By applying the procedure outlined in the Method and in Supporting Information we obtain two--particle wave function in an explicit form
\begin{strip}
\begin{align}
\begin{split}
     &\Psi_0(\textbf{r}_1,\textbf{r}_2) = \frac{2(t+V)}{\sqrt{2D(D-U+K)}} \Psi_{cov}(\textbf{r}_1,\textbf{r}_2) 
     -\frac{1}{2}\sqrt{\frac{D-U+K}{2D}}\Psi_{ion}(\textbf{r}_1,\textbf{r}_2) \equiv \\ &\equiv  C\psi_{cov}(\textbf{r}_1,\textbf{r}_2) + I\psi_{ion}(\textbf{r}_1,\textbf{r}_2),
\end{split}
     \label{coeff_wav1}
\end{align}
\end{strip}
where the covalent ($\Psi_{cov}$) and ionic ($\Psi_{ion}$) parts are
\begin{align}
    &\Psi_{cov}(\textbf{r}_1,\textbf{r}_2) = \left[ 
    w_1(\textbf{r}_1)w_2(\textbf{r}_2)+w_1(\textbf{r}_2)w_2(\textbf{r}_1)\right] \notag \\
    &\times\left[
    \chi_{\uparrow}(\textbf{r}_1)\chi_{\downarrow}(\textbf{r}_2)-\chi_{\downarrow}(\textbf{r}_1)\chi_{\uparrow}(\textbf{r}_2)\right], 
    \label{wav1} \\
    &\Psi_{ion}(\textbf{r}_1,\textbf{r}_2) = \left[  
    w_1(\textbf{r}_1)w_1(\textbf{r}_2)+w_2(\textbf{r}_1)w_2(\textbf{r}_2)\right] \notag \\ &\times\left[  
    \chi_{\uparrow}(\textbf{r}_1)\chi_{\downarrow}(\textbf{r}_2)-\chi_{\downarrow}(\textbf{r}_1)\chi_{\uparrow}(\textbf{r}_2)\right], 
    \label{wav2}
\end{align}

\noindent
where $\{w_i(\textbf{r})\}_{i=1,2}$ are trial H\"uckel-Slater molecular orbitals and 
$\{\chi_{\sigma}(i)\}_{\sigma=\pm1}$ are spin functions. In above expression the microscopic parameters, $U$ and $K$, are the magnitudes of intraatomic and interatomic Coulomb repulsion, respectively, $t$ and $V$ are the magnitudes of hopping (Bloch integral) and so--called correlated hopping, respectively (for details see Methods) and 

\begin{align}
    D \equiv \sqrt{(U-K)^2+16(t+V)^2}.
    \label{D}
\end{align}

We note that the two--particle wave function has the Heitler--London form, except the 
coefficients contain \textbf{all} interparticle interaction terms and the single--particle wave functions are the molecular H\"uckel-Slater orbitals with an adjusted size in the 
resultant correlated state. This two--particle wave function may be rewritten in terms of
the original Slater orbitals as follows
\begin{strip}
\begin{align}
\begin{split}
     &\Psi_0(\textbf{r}_1,\textbf{r}_2) = \left(C\beta^2(1+\gamma^2) - 2\gamma I \beta^2\right)\phi^{at}_{cov}(\textbf{r}_1,\textbf{r}_2)  
     +\left( I\beta^2(1-\gamma^2) - 2\gamma C \beta^2 \right)\phi^{at}_{ion}(\textbf{r}_1,\textbf{r}_2) \\
     &\equiv \tilde{C}\phi^{at}_{cov} + \tilde{I}\phi^{at}_{ion},
     \end{split}
     \label{coeff_wav2}
\end{align}
\end{strip}

\noindent
where the coefficients $\beta$ and $\gamma$ are defined through the relation

\begin{align}
    w_{i\sigma}(\textbf{r}) = \beta[\psi_{i\sigma}(\textbf{r})-\gamma
    \psi_{j\sigma}(\textbf{r})], 
    \label{wannier}
\end{align}

\noindent
where $i$ and $j$ label the atoms, $\sigma \equiv \pm 1 \equiv \uparrow, 
\downarrow$ is the electron spin quantum number. Note that 
$\braket{w_i(\textbf{r})|w_j(\textbf{r})} = \delta_{ij}$, $\beta$ and $\gamma$ are the mixing 
coefficients of the neighboring Slater orbitals ($\psi_{i\sigma} \equiv 
\psi(\textbf{r}-\textbf{R}_i)\chi_{\sigma}(i) \equiv 
(\sqrt{\alpha^3/\pi})exp(-\alpha|\textbf{r}-\textbf{R}_i|)\chi_{\sigma}(i)$), (in which 
$\alpha^{-1}$ is the size of the orbital). The functions $\phi^{at}_{cov}(\textbf{r}_1,\textbf{r}_2)$ and 
$\phi^{at}_{ion}(\textbf{r}_1,\textbf{r}_2)$ have the same form as \eqref{wav1} and \eqref{wav2}, respectively, 
except for the replacement $w_{i\sigma}(\textbf{r}) \rightarrow 
\psi_{i\sigma}(\textbf{r})$.

We define the effective atomic contribution $\tilde{A}$ at given interatomic distance $R$ as

                                                                                                 \begin{align}
   lim_{\gamma \rightarrow 0} \tilde{C} =C\beta^2 \equiv \tilde{A}, 
\end{align}                                           

\noindent
i.e., regard it as the weight of atomic part at the same $R$. Formally, the true covalency, ionicity, and atomicity are defined for given $R$ as

\begin{align}
    &\text{covalency: } \gamma_{cov} \equiv \frac{|\tilde{C}|^2-|\tilde{A}|^2}{|\tilde{C}|^2+|\tilde{I}|^2}, \label{cov}\\
    &\text{ionicity: }\gamma_{ion} \equiv 
    \frac{|\tilde{I}|^2}{|\tilde{C}|^2+|\tilde{I}|^2}, \label{ion}\\ 
    &\text{atomicity: }\gamma_{at} \equiv
    \frac{|\tilde{A}|^2}{|\tilde{C}|^2+|\tilde{I}|^2}. \label{atom}
\end{align}

The original motivation for introducing the concept of atomicity is as follows: It may seem natural to define covalency as $|C|^2/(|C|^2+|I|^2)$ and ionicity as $|I|^2/(|C|^2+|I|^2)$. However, such a choice leads directly to unphysical features (see Fig. \ref{fig:comp} in Supporting Information). Namely, $|C|^2/(|C|^2+|I|^2)$ reaches its maximal value of unity in the limit of separate atoms ($R\rightarrow \infty$). This is also the limit when electrons are entirely localized on their parent atoms and become distinguishable in the quantum mechanical sense. This is the limit which we regard as that atomicity equal to unity and vanishing true covalency. This type of argument is also the reason of subtracting the probability $\tilde{A}^2$ from $\tilde{C}^2$, not the corresponding wave--function amplitudes. The quantities \eqref{cov}--\eqref{atom} are proved next to be useful and of crucial importance.

To demonstrate the fundamental meaning of the introduced quantities we relate them to the criteria of Mott\cite{Mott} and Hubbard\cite{Hubbard1964} for onsets of localized (atomic) behavior. This is because the evolution of molecular \ch{H2} (electron--paired) state into individual separate singly occupied (atomic) states is regarded as equivalent to the Mott--Hubbard localization (for recent related discussion in different context see e.g. \cite{JPCLett2022}). Namely, we define Mott and Hubbard onset criteria as 

\begin{align}
    \frac{2|t+V|}{U-K} = 1, \text{      and      } n^{1/d}_c \alpha^{-1} \equiv \frac{1}{\alpha_0 R_{Mott}} \simeq 0.5, 
    \label{Hubbard-Mott}
\end{align}

\begin{figure}[t]
    \centering
    \includegraphics[width=0.5\textwidth]{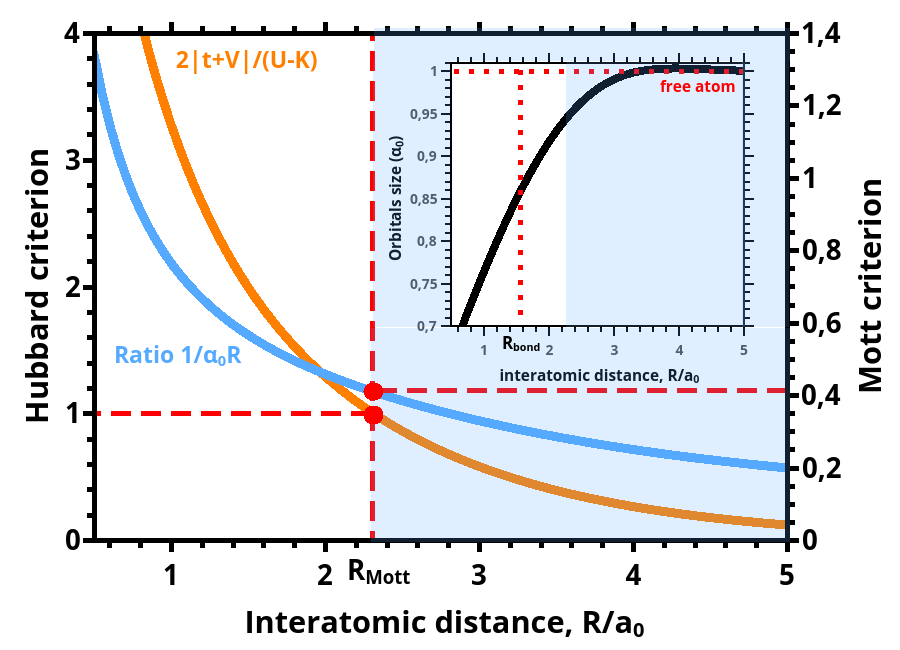}
    \caption{Hubbard (orange) and Mott (blue) characteristics of atomicity vs. interatomic distance $R$ (dashed horizontal lines). The dots mark the points corresponding to Hubbard and Mott criteria. The vertical dotted line marks the onset of \textit{Mottness} at $R_{Mott}$. The inset: $R$ dependence of the orbital size of the renormalized atomic wave functions composing the molecular (Wannier) single-particle states. The dotted line marks the equilibrium distance $R_{bond}$.}
    \label{fig:localization}
\end{figure}
\noindent
respectively, where $\alpha_0$ is the readjusted inverse orbital size, here at $R = R_{Mott}$. The first of them implies that for $R=R_{Mott}$ the kinetic (hopping) energy is equal to the correlation energy, i.e., for $R<R_{Mott}$ the ratio is greater than unity, whereas for $R>R_{Mott}$ it is smaller than unity and reduces quite rapidly to zero with increasing $R$ beyond $R_{Mott}$. In other words, the kinetic energy dominates in the former case and enhances hopping electrons to resonate strongly between the sites, whereas the electrons become gradually frozen as $R$ increases beyond $R_{Mott}$. 
On the other hand, the Mott criterion expresses the onset of localization in terms of the renormalized single--particle wave function size at the localization threshold. Namely, the threshold is reached when the diameter of the orbital in the correlated state ($2\alpha_0^{-1}$) is equal to the interorbital distance ($R=R_{Mott}$). Semiclassically, it means that the collective character is established when the orbitals start overlapping.
\begin{figure}[t]
    \centering
    \includegraphics[width=0.48\textwidth]{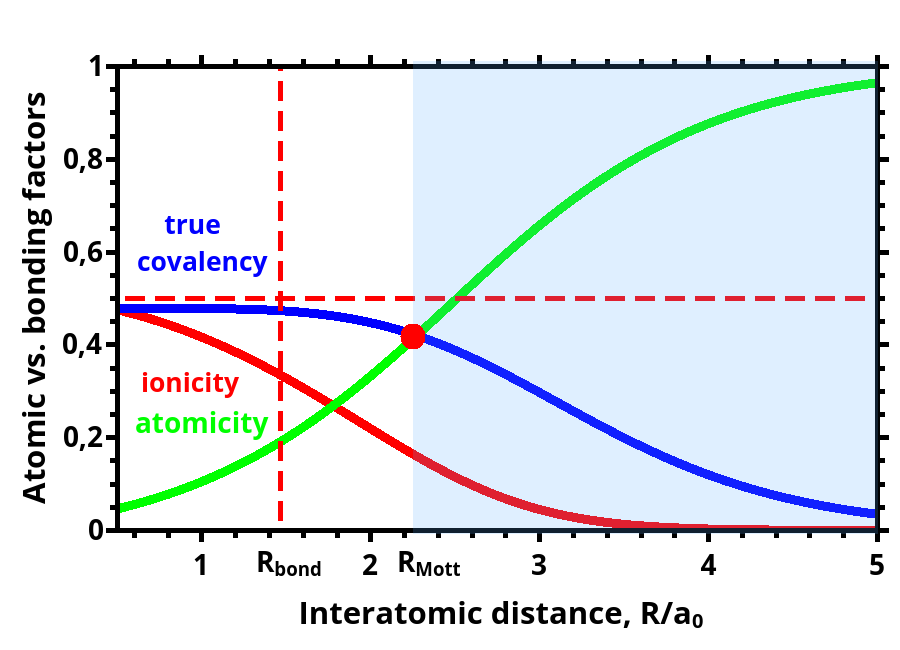}
    \caption{Intrinsic properties of the chemical bond: atomicity (green), true covalency (blue), and ionicity (red), all as a function of interatomic distance $R$. They represent the relative weights in the total two-particle wave function. In the $R\rightarrow 0$ limit the atomicity practically disappears and is the only contribution in the separate--atom limit $R \rightarrow \infty$. The solid circle defines the onset of localization effects (\textit{Mottness}) due to interelectronic correlations. If the atomicity is disregarded, the covalency exhibits a drastic nonphysical behavior with increasing $R > R_{bond}$. The figure illustrates a systematic evolution of molecular states into separate atoms and vice versa, formation of molecular states out of separate atoms. The Slater states have a renormalized size $\alpha^{-1}\leq\alpha_B$.}
    \label{fig:covionat}
\end{figure}
To visualize this formal reasoning we have plotted in Fig. \ref{fig:localization} the left parts of both \eqref{Hubbard-Mott} as a function of $R$, as well as have marked by red points their values at $R_{Mott}$. The blue shaded area may be called \textit{the Mottness regime} \cite{PHILLIPS20061634}. Inset shows the corresponding dependence of the renormalized size of the Slater orbital in the correlated state, with the dotted vertical line marking the equilibrium bond length $R_{bond} \simeq 1.43a_0$. Note that the orbital size for $R>R_{Mott}$ approaches rapidly to the free--atom values $a_0$.
\begin{figure*}[htbp]
    \centering
    \includegraphics[width=1\textwidth]{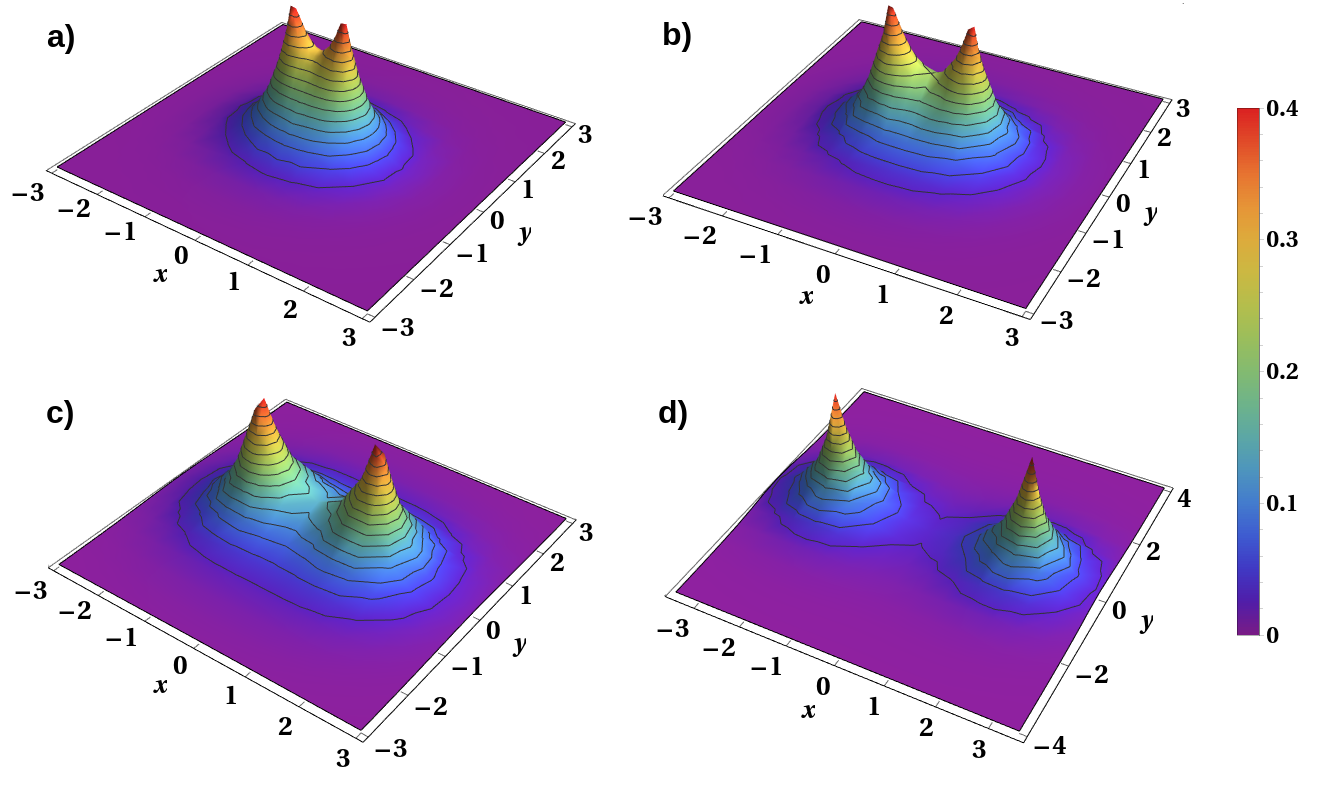}
    \caption{Electron density $n_{\sigma}(\textbf{r})$ for different interatomic distance (a) $R=1a_0^{-1}$, (b) $R=1.43a_0^{-1}$, (c) $R=2.3a_0^{-1}$, (d) $R=4a_0^{-1}$. The parts centered at nuclei are practically disjoint for $R \geq 5a_0$, illustrating the robustness of atomic behavior in that situation. This density contains also the double-occupancy (ionicity) contribution which becomes rapidly negligible with increasing $R$ beyond $R_{Mott}$.}
    \label{fig:density}
\end{figure*}
The explicit connection of the above onset to the true covalency ($\gamma_{cov}$), ionicity ($\gamma_{ion}$), and atomicity ($\gamma_{at}$) is visualized in Fig. \ref{fig:covionat}, where the $R$ dependence of those quantities is drawn. Remarkably, at the distance $R_{Mott}$ the true covalency and atomicity acquire the same value, so the point $R_{Mott}$ is a crossover point from true--covalency dominated to atomicity (\textit{Mottness}) regime. Furthermore, $\gamma_{cov}$ is predominant for $R<R_{Mott}$, whereas $\gamma_{at}$ is for $R>R_{Mott}$, as it should. Additionally, the covalency and ionicity (atom double occupancy) coincide as $R \rightarrow 0$, whereas then $\gamma_{at} \rightarrow 0$. Both $\gamma_{cov}$ and $\gamma_{ion}$ disappear in the atomic limit ($R>>R_{Mott}$), where $\gamma_{at} \rightarrow 1$. The results presented in Fig. \ref{fig:covionat} illustrate one the central findings of the present work. The principal characteristics are detailed further in the Tables \ref{tab:mott} and \ref{tab:values}. In Table \ref{tab:mott} we list the discussed factors of Mottness onset to show that they are mutually consistent. This agreement leads to the conclusion that the introduced entities \eqref{cov}--\eqref{atom} are not only relevant for the description of the Mott--Hubbard localization in condensed-matter (extended) systems, but also appear as a crucial incipient feature in molecular systems. We stress, this was possible only by introducing two-particle wave function as the proper characteristic of single bond which, after all, is composed of electron pairs. 
\begin{table}[t]
\centering
\caption{Equivalent characteristics of the atomicity onset threshold (for details see main text).}
\begin{tabular}{|c|c|}
\hline
\textbf{Characteristic}   & \textbf{Value} \\ \hline
Mott criterion            & 0.42           \\ \hline
Hubbard criterion        & 1              \\ \hline
Covalent-atomic crossover & 2.285$a_0$      \\ \hline
$R_{Mott}$                &   2.279$a_0$               \\ \hline
\end{tabular}
\label{tab:mott}
\end{table}

\begin{table}[htbp]
\resizebox{0.48\textwidth}{!}{
\centering
\caption{Particles density at the mid point, inverse orbital size $\alpha_0$, and mixing coefficients $\beta$ and $\gamma$, all versus $R$. The particle density $n(0,0)$ illustrates the gradually vanishing electron density in the region between the atoms as $R$ increases beyond $R_{Mott}$.}
\begin{tabular}{|c|c|c|c|c|}
\hline
R ($a_0$) & $n(0,0)$  & $\alpha_0$ & $\beta$ & $\gamma$ \\ \hline
1     & 0.334 & 1.30751                & 1.1386                 & 0.47811                \\ \hline
1.43  & 0.267 & 1.19838                & 1.0854                 & 0.38877                \\ \hline
2.3   & 0.142 & 1.05428                & 1.0338                 & 0.25358                \\ \hline
4     & 0.023 & 0.998601               & 1.0012                 & 0.048640               \\ \hline
\end{tabular}
\label{tab:values}}
\end{table}

So far, our discussion was based on the wave--function mechanics. In the 
remaining part we reformulate the analysis directly in the second--quantization language which will allow us to provide the physical interpretation of the bond in 
terms of particle densities. Namely, to amplify our multiparticle bond description we return to the particle language and display in Fig. \ref{fig:density} several panels composed of electron density in the $(x,y)$ plane, with the protons distant by $R/a_0 = 1,$ $1.43$($R_{bond}$), $2.3$($R_{Mott}$), and $4$ (profiles a) - d)), respectively. The density is defined as

\begin{align}
    n_{\sigma}(\textbf{r}) = \bra{\psi_G} \hat{\psi}^{\dagger}_{\sigma}(\textbf{r})\hat{\psi}_{\sigma}(\textbf{r}) \ket{\psi_G}, 
    \label{dens}
\end{align}

\noindent
where $\ket{\psi_G}$ is the lowest spin-singlet eigenstate and $\hat{\psi}_{\sigma}(\textbf{r})$ is the 
field operator. One should note that this density when integrated and summed over spin directions 
($\sigma\pm 1$) is equal to the total number of particles ($N_e = 2$). Obviously, $n_{\uparrow}(\textbf{r}) = n_{\downarrow}(\textbf{r}) \equiv n(\textbf{r})$/2, where $n(\textbf{r})$ is the total density. More importantly, this quantity 
provides the physical density, in contrast to the probability density 
$|\psi_G(\textbf{r}_1,\textbf{r}_2)|^2$. This distinctive feature of $n(\textbf{r})$ shows that the 
density diminishes in the region between the atoms to zero relatively fast with the increasing $R$ above $R_{Mott}$. To substantiate
the last statement we have listed in Tab. \ref{tab:values} the density value $n(0,0)$ in the middle point between 
the proton positions. For the sake of comparison, we have also added there the inverse orbital size, as well as 
the mixing coefficients in the wave function $w_i(\textbf{r}) = 
w(\textbf{r}-\textbf{R}_i)$ to show, that indeed both two-- and single--particle characteristics merge into 
their atomic correspondants as $R$ increases beyond $R_{Mott}$. Note that whereas $\gamma$ expresses the decreasing Pauling covalency \cite{pauling1960nature}, $n(0,0)$ describes the diminishing true covalency.

Analogously, we can define the concentration of the electrons in ionic configuration as follows:

\begin{align}
    &n_{ion}(\textbf{r}) \equiv \bra{\psi_G} \hat{n}_{\sigma}(\textbf{r})
    \hat{n}_{\bar{\sigma}}(\textbf{r}) \ket{\psi_G} 
    \label{ion_dens}
\end{align}

\noindent
with $\hat{n}_{\sigma}(\textbf{r}) \equiv \hat{\psi}^{\dagger}_{\sigma}(\textbf{r})\hat{\psi}_{\bar{\sigma}}(\textbf{r})$ and $\bar{\sigma} \equiv - \sigma$.
In other words, $n_{ion}(\textbf{r})$ expresses the density of local spin-singlet pairs (double-site occupancy). Profiles of $n_{ion}(\textbf{r})$ are not presented explicitly as they are contained in the equivalent form as the second part of the wave function \eqref{coeff_wav2}. 

Finally, along with the definitions \eqref{dens} and \eqref{ion_dens} of local particle densities, we can define nonlocal density of covalent electrons in the following manner

\begin{align}
    n_{cov}(\textbf{r}_1,\textbf{r}_2) = \bra{\psi_G} \hat{n}_{\sigma}(\textbf{r}_1) \hat{n}_{\bar{\sigma}}(\textbf{r}_2)\ket{\psi_G}.
    \label{cov_dens}
\end{align}

\noindent
This expression for correlation function completes our description in both the first and second--quantization schemes. The expressions \eqref{dens}--\eqref{cov_dens} may be useful in the situations with more involved orbitals.

To summarize, our fairly complete analysis of single chemical bond is based entirely on the multiparticle description, both in the first and second--quantization schemes. Both of the descriptions are equivalent, but within the second of them it is possible to relate it directly to the particle language. In general, the approach may represent a starting point to a precise experimental resolution of the bond factors, as well as to quantify  their specific features in more complex situations. Also, the method bridges the atomic and molecular aspects of the chemical bond in precise multiparticle categories. Analogous analysis may be carried out for the resolution of covalent bonds evolution into ionic bonds and vice versa \cite{Hendzel2022}.


\section{METHOD}

Our analysis starts from full form of Hamiltonian in second quantization, with all interaction terms between electrons on the lowest orbitals, i.e.,

\begin{strip}
\begin{align}
     \mathcal{\hat{H}} = & \epsilon_a \sum_i \hat{n}_{i\sigma} 
    + {\sum_{ij\sigma}}' t_{ij}\,\hat{a}^{\dag}_{i\sigma} \,\hat{a}_{j\sigma} + U \sum_{i} \hat{n}_{i\uparrow}\,\hat{n}_{i\downarrow} + \frac{1}{2} {\sum_{ij}}' K_{ij}\hat{n}_{i\sigma}\,\hat{n}_{j\sigma'}
      \nonumber \\ 
    & - \frac{1}{2}{\sum_{ij}}' J^H_{ij}  \left(\hat{\textbf{S}}_i \cdot \hat{\textbf{S}}_j-\frac{1}{4}
    \hat{n}_i\hat{n}_j\right) + \frac{1}{2} {\sum_{ij}}' {J}'_{ij}
    (\hat{a}^{\dagger}_{i\uparrow}\hat{a}^{\dagger}_{i\downarrow}\hat{a}_{j\downarrow}\hat{a}_{j\uparrow} + \mathrm{H.c.})  \nonumber \\
    &+ \frac{1}{2} {\sum_{ij}}' V_{ij} (\hat{n}_{i\sigma}+\hat{n}_{j\sigma})(\hat{a}^{\dagger}_{i\bar{\sigma}}\hat{a}_{j\bar{\sigma}}+ \mathrm{H.c.}) + \mathcal{H}_{\text{ion-ion}},
    \label{Hamiltonian_eq}
\end{align}
\end{strip}

\noindent
\begin{figure}[ht]
    \centering
    \includegraphics[width=0.46\textwidth]{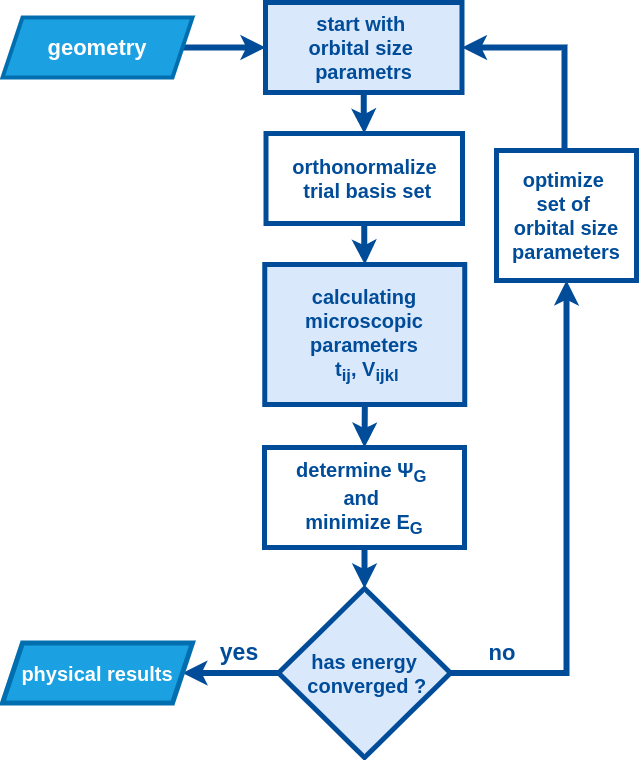}
    \caption{Flowchart of the EDABI method. The method is initialized by selection of a trial single-particle basis of wave functions \eqref{wannier}, and subsequent diagonalization of the many-particle Hamiltonian \eqref{Hamiltonian_eq}. Optimization of the single-particle-state size leads to an explicit determination of the trial-wavefunction parameters, microscopic interaction and hopping parameters as well as ground-state energy, and explicit form of the many-particle wavefunction, all in the correlated interacting state for a given interatomic distance.}
    \label{fig:flowchart}
\end{figure}
where $H.c$. denotes the Hermitian conjugation, $\hat{a}_{i\sigma}$ 
($\hat{a}^\dagger_{i\sigma}$) are fermionic annihilation (creation) operators for state $i$ 
and spin $\sigma$, $\hat{n}_{i\sigma} \equiv \hat{a}^\dagger_{i\sigma} \hat{a}_{i\sigma}$, and
$\hat{n}_i \equiv \hat{n}_{i\uparrow} + \hat{n}_{i\downarrow} \equiv
\hat{n}_{i\sigma}+ \hat{n}_{i\bar{\sigma}}$. The spin operators are defined as $\hat{S}_i 
\equiv \frac{1}{2} \sum_{\alpha\beta} \hat{a}^\dagger_{i\alpha} \sigma_i^{\alpha\beta} 
\hat{a}_{i\beta}$ with $\{\sigma_i^{ab}\}$ representing Pauli matrices. The Hamiltonian contains the 
atomic and hopping parts ($\propto \epsilon_a$ and $t_{ij}$, respectively), the so-called 
Hubbard term $\propto U$,  representing the  intra-atomic interaction between
the particles on the same atomic site \emph{i} with opposite spins, the direct intersite 
Coulomb interaction $\propto K_{ij}$, Heisenberg exchange $\propto J^H_{ij}$, and the 
two-particle and the correlated hopping and intersite Coulomb terms ($\propto J_{ij}^\prime$ and $V_{ij}$, 
respectively). The last term describes the ion-ion Coulomb interaction which is adopted here 
in its classical form. The microscopic parameters ($\epsilon_a$, $t_{12} \equiv t$, $U$, $K_{12} = K$, $J^H_{12} = {J}'_{12}$ and $V_{12} = V$) are all calculated explicitly in the resultant correlated state by readjusting the single-particle wave function size contained in their expressions (for their analytic expressions see Supporting Information). The primed summations are taken for $i\neq j$. The evolution of the new introduced quantities: atomicity, true covalency, and ionicity, is analyzed in detail as a function of interatomic distance.

The Hamiltonian \eqref{Hamiltonian_eq} was determined by defining first the field operators $\hat{\psi}_{\sigma}(\textbf{r})$ and $\hat{\psi^{\dagger}}_{\sigma}(\textbf{r})$, i.e., 

\begin{align}
\begin{cases}
    \hat{\psi}_{\sigma}(\textbf{r}) = \sum_{i\sigma}w_i(\textbf{r})\chi_{\sigma}(i)\hat{a}_{i\sigma}, \\
    \hat{\psi^{\dagger}}_{\sigma}(\textbf{r}) = \sum_{i\sigma}w_i(\textbf{r})\chi_{\sigma}^{\dagger}(i)\hat{a}^{\dagger}_{i\sigma},
    \end{cases}
    \label{field_operator}
\end{align}

\noindent
where $\hat{a}_{i\sigma}$ ($\hat{a}^{\dagger}_{i\sigma}$) are the annihilation (creation) operators
of the single-particle states $w_i(\textbf{r})\chi_{\sigma}(i)$ on atom $i=1,2$ with spin $\sigma$. Note that the single--particle basis is composed of the H\"uckel-Slater orbitals only; this restriction represents the only approximation here; all the remaining analysis and results are exact within these limitations. Those operators, in turn lead to the expression \eqref{Hamiltonian_eq} of $\hat{\mathcal{H}}$ in a standard manner. To close the formal methodological part, the two-particle spin-singlet wave function is defined as \cite{Robertson}

\begin{align}
    \psi_G(\textbf{r}_1,\textbf{r}_2) = \frac{1}{\sqrt{2}} \bra{0} \hat{\psi}_1(\textbf{r}_1)
    \hat{\psi}_2(\textbf{r}_2)\ket{\psi_G}, 
\end{align}

\noindent
where $\ket{0}$ is the universal vacuum state for particles and $\ket{\psi_G}$ is the 
ground state, both in Fock space. This relation provides equivalence of description both 
in terms of two-particle wave function and the second-quantization language.

In Fig. \ref{fig:flowchart} we provide the flowchart of the numerical part of our analysis which concerns mainly the determination of the orbital size $\alpha^{-1}$ in the interacting (correlated) state and, in effect, of the optimal ground state energy and all other bonding--state characteristics. 

\begin{acknowledgement}
This work was supported by Grants OPUS No.~UMO--2018/29/B\\/ST3/02646 and 
No.~UMO--2021/41/B/ST3/04\\070 from Narodowe Centrum Nauki. Discussions with our colleague from 
Theoretical Chemistry Department, Prof. Ewa Broc\l{}awik and dr hab.~Mariusz Rado\'n from 
Inorganic Chemistry Department of the Jagiellonian University are gratefully acknowledged. The authors are also grateful to Prof. Jurgen M. Honig from Purdue University for his critical reading of the manuscript.
\end{acknowledgement}

\bibliography{bibliography}

\clearpage
\newpage
\begin{suppinfo}

The formal solution of Hamiltonian \eqref{Hamiltonian_eq} is carried out explicitly by selecting a trial basis in the Fock space for this two orbital systems. In the case with $N=4$ spinorbitals and $N_e=2$ electrons we have $\binom{N}{N_e} = \binom{4}{2} = 6$ states in the occupation number representation. They are
\begin{align}
\begin{cases}
    &\ket{1} = \hat{a}^{\dagger}_{1\uparrow}\hat{a}^{\dagger}_{2\uparrow} \ket{0},\\
    &\ket{2} = \hat{a}^{\dagger}_{1\downarrow}\hat{a}^{\dagger}_{2\downarrow}\ket{0},\\
    &\ket{3} = \frac{1}{\sqrt{2}}(
    \hat{a}^{\dagger}_{1\uparrow}\hat{a}^{\dagger}_{2\downarrow}+\hat{a}^{\dagger}_{1\downarrow}\hat{a}^{\dagger}_{2\uparrow})\ket{0},\\
    &\ket{4} = \frac{1}{\sqrt{2}}(
    \hat{a}^{\dagger}_{1\uparrow}\hat{a}^{\dagger}_{2\downarrow}-\hat{a}^{\dagger}_{1\downarrow}\hat{a}^{\dagger}_{2\uparrow})\ket{0},\\
    &\ket{5} = \frac{1}{\sqrt{2}}(
    \hat{a}^{\dagger}_{1\uparrow}\hat{a}^{\dagger}_{1\downarrow}+\hat{a}^{\dagger}_{2\downarrow}\hat{a}^{\dagger}_{2\uparrow})\ket{0},\\
    &\ket{6} = \frac{1}{\sqrt{2}}(
    \hat{a}^{\dagger}_{1\uparrow}\hat{a}^{\dagger}_{1\downarrow}-\hat{a}^{\dagger}_{2\downarrow}\hat{a}^{\dagger}_{2\uparrow})\ket{0}.
    \end{cases}
\end{align}\

\noindent
The first three are the spin--triplet states, whereas the next three are the spin--singlets and the mixture of those will result in the ground state. By calculating the Hamiltonian matrix $\bra{i}\hat{\mathcal{H}} \ket{j}$ we obtain a $6\times6$ matrix which can be diagonalized analytically (see Ref.~\cite{JS2000}). In the case of interest to us case we list only the ground state eigenvalue and corresponding ground states which are 

\begin{align}
    \begin{cases}
        E \equiv \lambda_5 =  2\epsilon + \frac{1}{2}(K+U) + J - \frac{1}{2} D, \\ 
        \ket{\psi_G} = \frac{1}{[D(D - U + K)]^{\frac{1}{2}}} [(4(t+V))\ket{4}  \\
        - (D - U + K)\ket{5}], 
    \end{cases}
\end{align}

\noindent
where parameters are defined in the main text. Their explicit expressions are

\begin{align}
\resizebox{0.46\textwidth}{!}{
    $\begin{cases}
        \epsilon_a = \beta^2(1 + \gamma^2)\epsilon_a' - 2\beta^2\gamma t' \\
        t = t_{12} = \beta^2(1 + \gamma^2)t' - 2\beta^2\gamma\epsilon_a ',\\
        U = \beta^4[(1 + \gamma^4)U' + 2\gamma^2K' 
        -4\gamma(1+ \gamma ')V' + 4\gamma^2 J'],\\
        K = K_{12} = \beta^4[(1 + \gamma^2)K' + 2\gamma^2U'
        -4\gamma(1+ \gamma ')V' \\+ 4\gamma^2 J'],\\
        V = V_{12} = \beta^4[ \gamma^2)K' + 2\gamma^2U' 
        -4\gamma(1+ \gamma ')V' \\+ (1+ \gamma^2) J'\\
        J^{H} \equiv J^H_{12} = {J}'_{12} = \beta^4[-\gamma(1 + \gamma^2)U' \\
        -\gamma(1 + \gamma^2)K' 
    + (1 + 6\gamma^2 + \gamma^4)V'\\
    -2\gamma(1 + \gamma^2J')].
    \end{cases}$}
\end{align}

\noindent
where $\beta$ and $\gamma$ are mixing parameters and ${\epsilon}'$, ${t}'$, ${U}'$, ${K}'$, ${J}'$, ${V}'$ are analogous microscopic parameters calculating for single-particle wave functions. They can be expressed as a function of interatomic distance $R$ and inverse orbital
size $\alpha_0$

\begin{align}
\resizebox{0.5\textwidth}{!}{
    $\begin{cases}
        {\epsilon}'_a =\alpha_0^2-2\alpha_0-\frac{2}{R}+2( \alpha_0+\frac{1}{R} )\text{exp}(-2\alpha_0 R), \\
        {t}' = \alpha_0^2\text{exp}(-\alpha_0 R) [1+\alpha_0 R +\frac{1}{3}\alpha_0^2 R^2] \\
        -4\alpha_0 (1+\alpha_0 R)\text{exp}(-\alpha_0 R), \\ 
        {U}' = \frac{5}{4}\alpha_0, \\ 
        {K}' = \frac{2}{R}-\alpha_0 \text{ exp}(-\alpha_0 R) [ \frac{2}{\alpha_0 R}+\frac{3}{2}\alpha_0 R \\+ 
        \frac{1}{3}(\alpha_0 R)^2 +\frac{11}{4} ],\\ 
        {V}' = \alpha_0 [ \text{exp}(-\alpha_0 R)(2\alpha_0 R +\frac{5}{8\alpha_0 R} +\frac{1}{4} )\\ -\frac{1}{4}(1+\frac{5}{2\alpha_0 R})\text{exp}(-3\alpha_0 R)],\\ 
        {J}' = \frac{12}{5R}[S^2C+S^2 \text{ ln}(\alpha_0 R)-2SS'E_i(-2\alpha_0 R)\\+(S')^2E_i(-4\alpha_0 R)]\\
        +\alpha_0 \text{ exp}(-2\alpha_0 R)[\frac{5}{4}-\frac{23}{10}\alpha_0 R -\frac{6}{5} \alpha_0^2R^2-\frac{2}{15}\alpha_0^3R^3],
    \end{cases}$}
    \end{align}

\noindent
with the Euler constant $C \simeq 0.57722$, and 

\begin{align}
\begin{cases}
    E_i(x)=-\int^\infty_x \frac{dt}{t}\text{ exp}(-t),\\
    S= \text{exp}(-\alpha R)\left(1+\alpha R +\frac{1}{3}\alpha^2R^2 \right),\\
    S'=\text{exp}(\alpha R)\left(1-\alpha R +\frac{1}{3}\alpha^2R^2 \right).
    \end{cases}
\end{align}

In effect, the determination of $\ket{\psi_G}$ requires the readjustment of the Slater-orbital size contained in the expression for H\"uckel-Wannier orbitals $w_i(\textbf{r})$, a procedure of which is schematically illustrated by the flowchart in Fig. \ref{fig:flowchart}.

The microscopic parameters obtained in this way are shown in Figs.~\ref{fig:param1} and~\ref{fig:param2}. Parenthetically, the mixing coefficients $\beta$ and $\gamma$ are expressed through the overlap integral $S \equiv\braket{\psi_1(\textbf{r})^{\dagger}\psi_2(\textbf{r})}$ 

\begin{align}
    \begin{cases}
        \beta = \frac{1}{\sqrt{2}}\sqrt{\frac{1+\sqrt{1+S^2}}{1-S^2}}, \\
        \gamma = \frac{S}{1+\sqrt{1-S^2}}. \\
    \end{cases}
\end{align}

\noindent
\begin{figure}[htbp]
    \centering
    \includegraphics[width=0.45\textwidth]{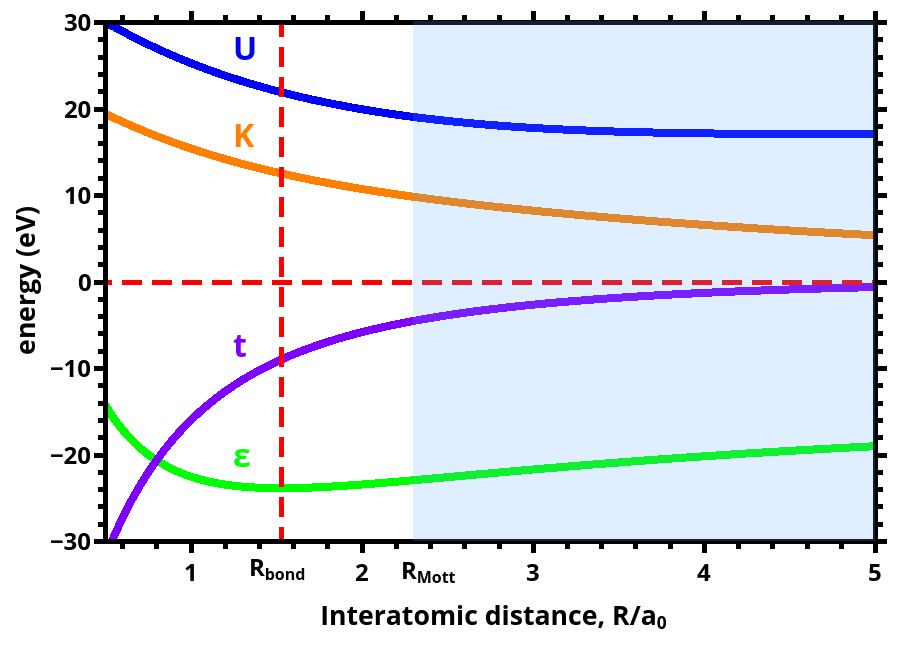}
    \caption{Microscopic parameters indicated versus interatomic distance $R$. The equilibrium bond length and the onset of \textit{Mottness}, $R_{Mott}$ are also marked.}
    \label{fig:param1}
\end{figure}
The whole procedure is closed once we calculate explicitly the ground-state two-particle wave function according to the general rule 

\begin{align}
    &\psi_G(\textbf{r}_1,\textbf{r}_2) = \frac{1}{2} \bra{0} \{ \hat{\psi}_{\uparrow}(\textbf{r}_1)\hat{\psi}_{\downarrow}(\textbf{r}_2) \notag \\&- \hat{\psi}_{\downarrow}(\textbf{r}_1)\hat{\psi}_{\uparrow}(\textbf{r}_2)\}\ket{\psi_G}
\end{align}
\begin{figure}[htbp]
    \centering
    \includegraphics[width=0.45\textwidth]{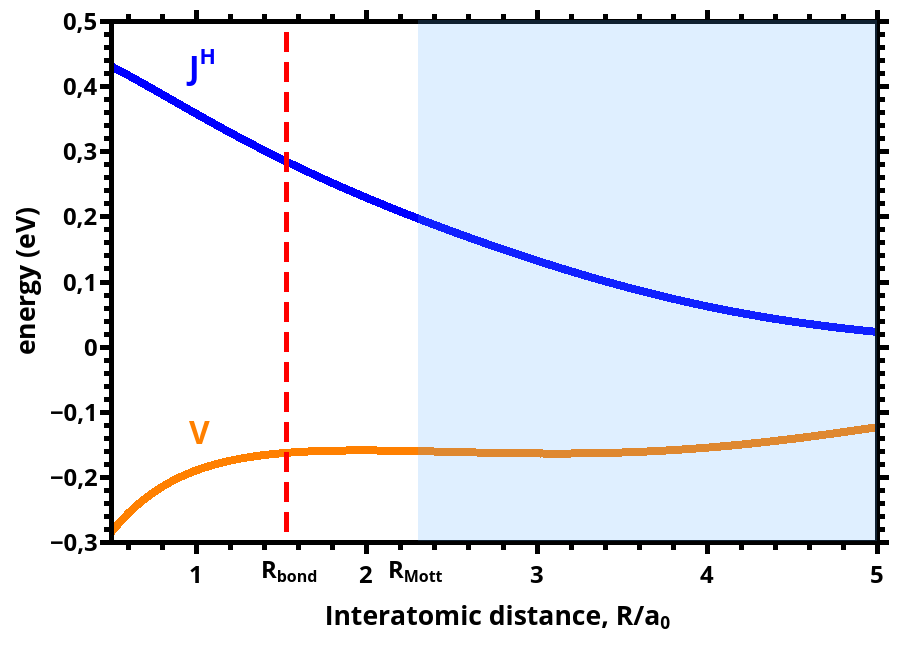}
    \caption{Microscopic parameters, $J^H$ and $V$ versus $R$ (same characteristics as in Fig.\eqref{fig:param1}).}
    \label{fig:param2}
\end{figure}
The final result is the expression \eqref{coeff_wav1}, which contains single-particle molecular wave functions $\{w_i(\textbf{r})\}_{i=1,2}$.

The problem one encounters at the outset is that if one examines the $R$ dependence of the coefficients $|C|^2/(|C|^2+|I|^2)$ and $|I|^2/(|C|^2+|I|^2)$; this has been plotted in Fig. \ref{fig:comp}. Namely, one spots that such naturally defined covalency exhibits a clear unphysical behavior in the limit $R \rightarrow \infty$, when the molecular states should reduce to the parent atomic Slater states. To avoid this basic deficiency the concept of atomicity was involved and discussed in detail in the main text.
\begin{figure}[htbp]
    \centering
    \includegraphics[width=0.45\textwidth]{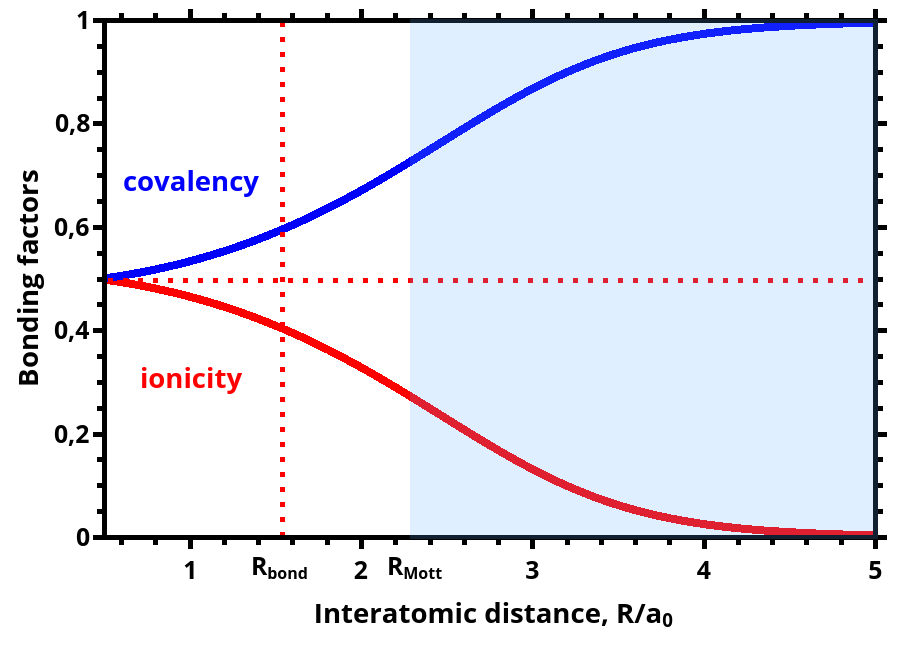}
    \caption{Two-particle covalency and ionicity for \ch{H2} molecule calculated as $|C|^2/(|C|^2+|I|^2)$ and $|I|^2/(|C|^2+|I|^2)$, respectively. These results are reinterpreted subsequently and the results, including atomicity, are displayed in Fig. \ref{fig:covionat} in the main text.}
    \label{fig:comp}
\end{figure}
Along with the explicit wave function expression (cf. Eqs.\eqref{coeff_wav1} and \eqref{coeff_wav2}) in main text), one can obtain the corresponding expression for the physical densities \eqref{dens}--\eqref{cov_dens}.
\end{suppinfo}


\end{document}